\documentclass[letter, longauth]{aa}

\usepackage{hyperref}
\usepackage{graphicx}
\usepackage{ulem}
\usepackage[varg]{txfonts}
\usepackage{lipsum}
\usepackage{lscape}    
\usepackage{placeins}           
\usepackage{xspace}
\usepackage{natbib}
\usepackage{siunitx} 
\usepackage{tabularx} 
\usepackage{textcomp} 
\usepackage[version=4]{mhchem} 
\usepackage[svgnames,dvipsnames]{xcolor} 
\usepackage{amsmath}  
\hypersetup{linkcolor=red,citecolor=blue,filecolor=cyan,urlcolor=magenta}

\renewcommand{\linenumbers}{}

\newcommand{\solar}[1]{\textit{#1}$_\odot$\xspace}

\newcommand{\micron}{$\mu$m\xspace}
\renewcommand{\arcsec}{$^{\prime\prime}$\xspace}

\newcommand{\cotwo}{\ce{CO2}\xspace}
\newcommand{\water}{\ce{H$_2$O}\xspace}
\newcommand{\thcotwo}{$^{13}$CO$_2$\xspace}
\newcommand{\twcotwo}{$^{12}$CO$_2$\xspace}

\begin{document}

   \title{Detection of CO$_2$ ice in the planetary nebula NGC~6302}



\author{%
    Charmi~Bhatt\inst{\ref{PAB},\ref{WesternSpace}} \and 
    Simon~W.~Cao\inst{\ref{PAB}} \and 
    Jan~Cami\inst{\ref{PAB},\ref{WesternSpace}}\and
    Nicholas Clark \inst{\ref{PAB}} \and
    Pascale Ehrenfreund \inst{\ref{Leiden},\ref{SPI}} \and
    Els Peeters \inst{\ref{PAB},\ref{WesternSpace}}\and
    Mikako Matsuura \inst{\ref{cardiff}} \and
    G.\ C.\ Sloan \inst{\ref{StScI}, \ref{carolina}} \and
    Harriet L. Dinerstein \inst{\ref{Utexas}} \and
    Patrick Kavanagh \inst{\ref{maynooth}} \and
    Kevin Volk \inst{\ref{StScI}} \and
    Isabel Aleman \inst{\ref{Aleman}} \and 
    Michael J. Barlow \inst{\ref{Barlow}} \and 
    Kay Justannont \inst{\ref{Chalmers}} \and  
    Kathleen E. Kraemer \inst{\ref{BostonCollege}} \and  
    Joel H. Kastner \inst{\ref{Kastner1},\ref{Kastner2},\ref{Kastner3}} \and  
    Francisca Kemper \inst{\ref{Kemper1}, \ref{Kemper2}, \ref{Kemper3}} \and  
    Hektor Monteiro \inst{\ref{cardiff}, \ref{Hektor}} \and  
    Raghvendra Sahai \inst{\ref{JPL}} \and  
    N.\ C.\ Sterling \inst{\ref{Sterling}} \and  
    Jeremy R. Walsh \inst{\ref{ESO}} \and  
    L. B. F. M. Waters \inst{\ref{Radboud}, \ref{SRON}} \and   
    Albert Zijlstra \inst{\ref{Manchester}}
    }
    \institute{
    Department of Physics and Astronomy, University of Western Ontario, London, Ontario, Canada\label{PAB} \and
    Institute for Earth and Space Exploration, University of Western Ontario, London, Ontario, Canada\label{WesternSpace} \and
    Leiden Observatory, Leiden University, PO Box 9513, 2300 RA Leiden, The Netherlands\label{Leiden} \and
    Space Policy Institute, George Washington University, 20052 Washington DC, USA\label{SPI} \and
    Cardiff Hub for Astrophysics Research and Technology (CHART), School of Physics and Astronomy, Cardiff University, The Parade, Cardiff CF24 3AA, UK\label{cardiff} \and 
    Space Telescope Science Institute, 3700 San Martin Drive, Baltimore, MD 21218, USA\label{StScI} \and
    Department of Physics and Astronomy, University of North Carolina, Chapel Hill, NC 27599-3255, USA\label{carolina} \and
    Department of Astronomy, University of Texas at Austin, Austin, TX 78712, USA\label{Utexas} \and
    Department of Physics, Maynooth University, Maynooth, County Kildare, Ireland\label{maynooth}
    \and
    Laborat\'{o}rio Nacional de Astrof\'{i}sica, Rua dos Estados Unidos, 154, Bairro das Na\c{c}\~{o}es, Itajub\'{a}, MG, 37504-365, Brazil\label{Aleman} \and
    Department of Physics and Astronomy, University College London, Gower Street, London WC1E 6BT, United Kingdom\label{Barlow} \and 
    Chalmers University of Technology, Onsala Space Observatory, 439 92 Onsala, Sweden\label{Chalmers} \and 
    Institute for Scientific Research, Boston College, 140 Commonwealth Avenue, Chestnut Hill, MA 02467, USA\label{BostonCollege} \and
    Center for Imaging Science, 
    Rochester Institute of Technology, Rochester NY 14623, USA\label{Kastner1} \and
    School of Physics and Astronomy,  Rochester Institute of Technology, Rochester NY 14623, USA\label{Kastner2} \and
    Laboratory for Multiwavelength Astrophysics,  Rochester Institute of Technology, Rochester NY 14623, USA\label{Kastner3} \and 
    Institut de Ci\`encies de l'Espai (ICE, CSIC), Can Magrans, s/n, E-08193 Cerdanyola del Vall\`es, Barcelona, Spain\label{Kemper1} \and 
    ICREA, Pg. Llu\'is Companys 23, E-08010 Barcelona, Spain \label{Kemper2} \and 
    Institut d'Estudis Espacials de Catalunya (IEEC), E-08860 Castelldefels, Barcelona, Spain\label{Kemper3} \and  
    Instituto de F\'isica e Qu\'imica, Universidade Federal de Itajub\'a, Av. BPS 1303 Pinheirinho, 37500-903 Itajub\'a, MG, Brazil\label{Hektor} \and
    Jet Propulsion Laboratory, 4800 Oak Grove Drive, California Institute of Technology,
    Pasadena, CA 91109, USA\label{JPL} \and 
    University of West Georgia, 1601 Maple Street, Carrollton, GA 30118, USA\label{Sterling} \and
    European Southern Observatory, Karl-Schwarzschild Strasse 2, D-85748 Garching, Germany\label{ESO} \and
    Department of Astrophysics/IMAPP, Radboud University, PO Box 19 9010, 6500 GL Nijmegen, The Netherlands\label{Radboud} \and
    SRON Netherlands Institute for Space Research, Niels Bohrweg 4, 2333 CA Leiden, The Netherlands\label{SRON} \and  
    Jodrell Bank Centre for Astrophysics, Department of Physics and Astronomy, The University of Manchester, Oxford Road,  Manchester M13 9PL, UK\label{Manchester} 
    }
   \date{}

   \abstract{Using JWST/MIRI observations, we report the detection  of \cotwo ice in the dusty torus of the planetary nebula NGC~6302, an environment generally considered hostile to fragile molecular species and ices due to intense UV irradiation. This detection accompanies cold (20-50~K) gas-phase \cotwo along the same sightlines. The ice absorption profile exhibits a double-peak profile, a characteristic of pure, crystalline \cotwo ice. The \cotwo gas-to-ice ratio is more than an order of magnitude higher than in young stellar objects, pointing to distinct ice formation or processing mechanisms in evolved stellar environments. This discovery demonstrates that the dusty torus provides sufficient shielding to harbour ice chemistry, and that ice-mediated surface reactions must be incorporated into chemical models of planetary nebulae.}


    \maketitle
\section{Introduction}

During the asymptotic giant branch (AGB) phase, low- to intermediate-mass ($\sim$\mbox{1--8~\solar{M}}) stars experience substantial mass loss \citep[$\sim$10$^{-8}$ to 10$^{-4}$ \solar{M} yr$^{-1}$; see][]{2018A&ARv..26....1H}, and as this stellar material expands outward it cools, enabling the formation of a wide variety of molecules and dust grains. As the central star evolves through the post-AGB phase, the effective temperature increases rapidly and when \textit{T}$_{\text{eff}}$ $\gtrsim$20,000K, intense UV radiation begins to ionize the circumstellar material, forming a planetary nebula (PN; plural: PNe) -- a generally hostile environment where photo-dissociation, photo-evaporation and energetic stellar outflows cause a dramatic decline in the molecular diversity compared to the AGB phase \citep{2022EPJWC.26500029A}. 
However, our understanding of the chemical pathways operating in PNe remains incomplete. JWST observations have only just begun revealing some of these key processes. Among the objects observed, the Butterfly Nebula (NGC 6302, a complex bipolar PN) has emerged as a particularly intriguing laboratory for investigating complex chemical pathways in PNe due to its extreme environment and surprisingly rich chemistry.

NGC~6302 is an oxygen-rich (O-rich) evolved object that was never a carbon star \citep{2011MNRAS.418..370W, 2025MNRAS.542.1287M}. Yet recent JWST/MIRI observations have revealed the presence of CH$_3^+$ \citep{Bhatt_2025b}, a key driver of organic chemistry \citep{smith_ion_1992, 2021FrASS...8..207H,   2023Natur.621...56B, 2025A&A...696A..99Z, 2025A&A...696A.100G} whose formation is enabled by the intense UV radiation from the central star. Moreover, the widespread presence of polycyclic aromatic hydrocarbons (PAHs) in this object \citep{2025MNRAS.542.1287M} is also puzzling, and large variations in its spectral features point to in-situ formation and/or processing (Clark et al., in prep). Both findings demonstrate that this extreme environment supports rich chemical processes, and NGC~6302 thus represents a particularly intriguing laboratory for investigating some of the complex chemical pathways in PNe. Here, we report another surprising discovery: the clear spectral signatures of cold \cotwo gas and the presence of \cotwo ice features, marking the first detection of \cotwo ice in a PNe.

\section{The torus of NGC~6302}
\label{sec:observations}

NGC~6302 exhibits bright east-west oriented bipolar lobes bisected by a massive dusty torus (oriented north-south; see Fig.~\ref{fig:HST}) that appears as a dark lane obscuring the central star at visible and near-infrared (IR) wavelengths \citep{2022ApJ...927..100K}. ALMA $^{12}$CO J$=$3--2 observations reveal that the torus is non-Keplerian and radially expanding at 8 km~s$^{-1}$, with a kinematical age of ~5000--7500 years \citep{2007A&A...473..207P, 2017AandA...597A..27S}. JWST mid-IR observations show that the torus is flared and warped, extending to a radius of 5.5\arcsec (5700 AU) at infrared wavelengths \citep{2025MNRAS.542.1287M}. The extinction in the torus at 5.9~\micron yields A$_\text{v}$ > 76 mag, indicating a (dust + gas) mass of 0.8--3\solar{M} and a hydrogen density of $n_H \sim 6 \times 10^{6}$ cm$^{-3}$\citep{2025MNRAS.542.1287M}. ISO-SWS spectra of NGC~6302 reveal the presence of crystalline silicates and carbonates \citep{2001A&A...372..165M, 2002AandA...394..679K}. 

This work utilizes JWST MIRI/MRS observations \citep{2015PASP..127..646W, 2023A&A...675A.111A} of NGC~6302 (program ID 1742; PI: M. Matsuura), covering the central star, torus, and innermost region of the bipolar lobes \citep[Fig.~\ref{fig:HST}; see][for more details]{2025MNRAS.542.1287M}. The data reduction was performed using JWST Calibration Pipeline version 1.16.1 and CRDS version 11.17.19 (see \citealt{2025MNRAS.542.1287M} and Clark et al., in prep, for details on the data reduction).

\section{Gas-phase \texorpdfstring{CO$_2$}{CO2} absorption}
\label{sec:gasCO2}

\begin{figure}
    \centering
    \resizebox{\hsize}{!}{\includegraphics{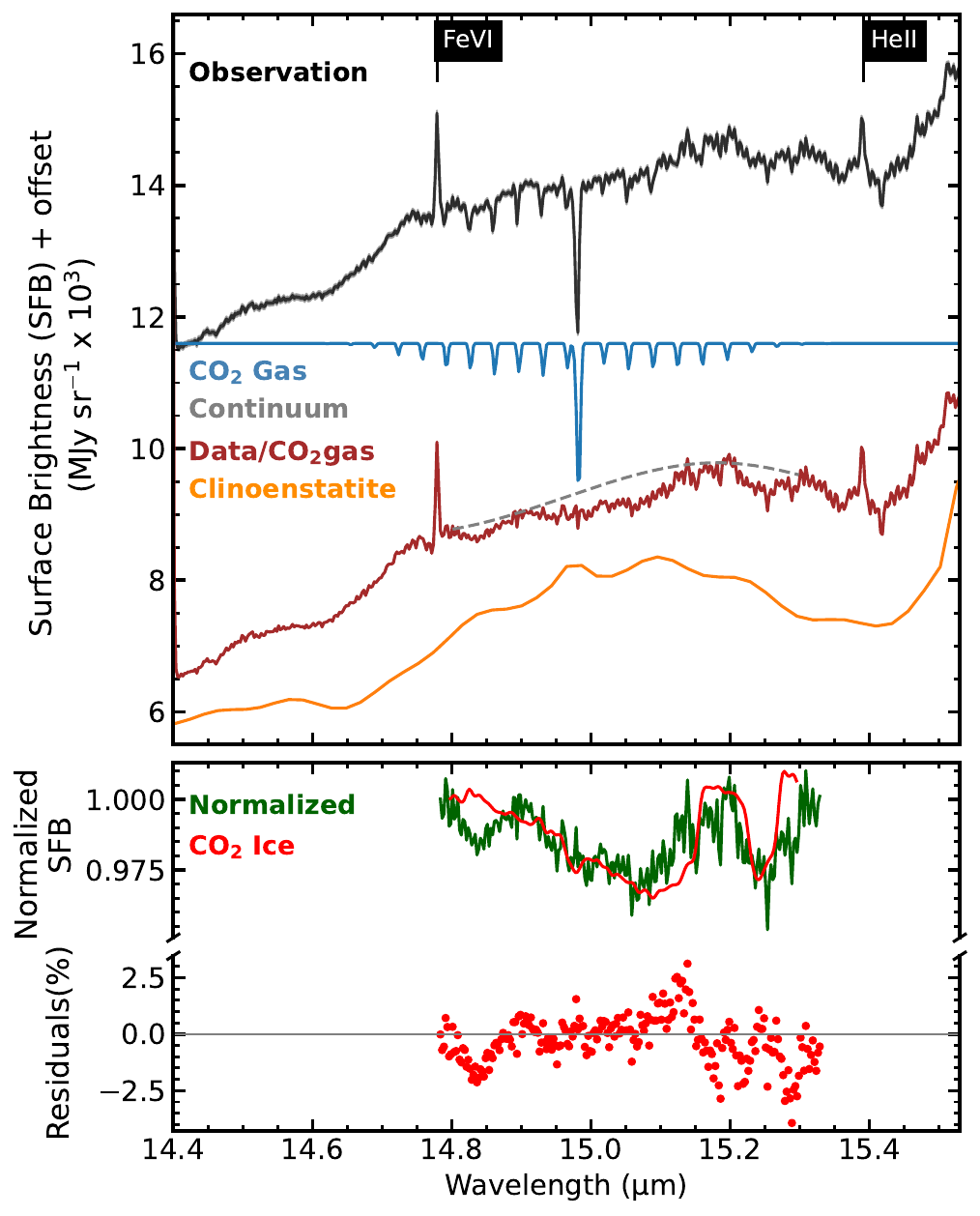}}
    \caption{Detection of \cotwo ice and gas. The black curve on top is the JWST/MIRI spectrum at the position indicated by the dot in Fig.~\ref {fig:HST}. Uncertainties on the surface brightness are $\sim \pm$ 40~MJy~sr$^{-1}$ . The best-fit gas-phase model ($T_{\rm gas}=30$~K and $\log N$=16.8~cm$^{-2}$) is shown in blue. We divided out this model, resulting in the brown curve. The orange curve shows a single-temperature model for clinoenstatite emission. The grey dashed line is the continuum we adopt to analyse the ice band. The green curve shows the  normalized spectrum using that continuum. The red curve is a laboratory spectrum for \cotwo ice (CO:CO$_2$ 2:1 ice mixture at 80~K); and the bottom panel shows the residuals between the normalized spectrum (green) and the ice model (red).}
    \label{fig:spectrum}
\end{figure}

The JWST/MIRI MRS spectra towards the northern part of the torus (see Fig.~\ref{fig:HST}) reveal clear absorption features in the 14.8--15.2\micron range corresponding to the $\nu_2$ bending mode of gas-phase \cotwo (see Fig.~\ref{fig:spectrum}). 

We calculated \cotwo model spectra as described in \citet{Cami:SF}, using the Carbon Dioxide Spectroscopic Database (CDSD) linelist for $^{12}$CO$_2$ \citep{2010hitr.confE...3T}. We adopted Gaussian intrinsic line profile with a line width $b$=1 km~s$^{-1}$ and smoothed and resampled the resulting spectra to match the JWST/MIRI resolution at 15~\micron. We used a model grid that spans temperatures of 10–100~K (in 10~K steps), and a logarithmic grid in column densities in the range of $\log~N$ = 14--19~cm$^{-2}$ (in steps of $\log N=0.1$). We also considered radial velocities in the range of $-$70 to $-$26~km~s$^{-1}$ (in 1~km~s$^{-1}$ steps) -- a slightly wider velocity range than found by ALMA observations \citep{2017AandA...597A..27S}. 

The \cotwo absorption occurs against a structured continuum. To determine this local continuum, we masked the narrow \cotwo absorption features and fitted a polynomial to the spectra. We then used this continuum to normalize the observed spectrum (see Fig.~\ref{fig:12CO2_and_13CO2fit}) and performed a least-squares minimization at each pixel to simultaneously determine the excitation temperature, column density, and radial
velocity of gas-phase \cotwo. We determined the 1$\sigma$ uncertainties on those best-fit parameters from the $\chi^2$ hypersurface. The LTE \cotwo slab models generally reproduce the absorption very well (see Fig.~\ref{fig:spectrum}). The \cotwo gas is cold everywhere we detect it, with temperatures in a narrow range of 20–50~K (with typical uncertainties of $\pm$10~K; Fig.~\ref{fig:Tmap}) for the majority of the spaxels. Column densities of this cold \cotwo gas increase smoothly by over an order of magnitude from the outer edge (log~N $=$ 15.2 $\pm$0.1 cm$^{-2}$) to the torus midplane (log~N $=$ 16.8 $\pm$0.1 cm$^{-2}$; Fig.~\ref{fig:HST}). For most pixels, the derived radial velocities, in the barycentric reference frame, range from $-$55 to $-$30 km~s$^{-1}$ (with typical uncertainties of $\pm$3 km~s$^{-1}$ ), similar to the range ($-$40 to $-$30 km~s$^{-1}$) found using ALMA $^{12}$CO J$=$3--2 observations \citep{2017AandA...597A..27S}. We also detect much weaker spectral features of gas-phase \thcotwo (see Appendix~\ref{app:12and13CO2}).

\section{Detection of \texorpdfstring{CO$_2$}{CO2} ice}
\label{sec:ices}

\begin{figure}[t]
    \centering
\includegraphics[width=\linewidth]{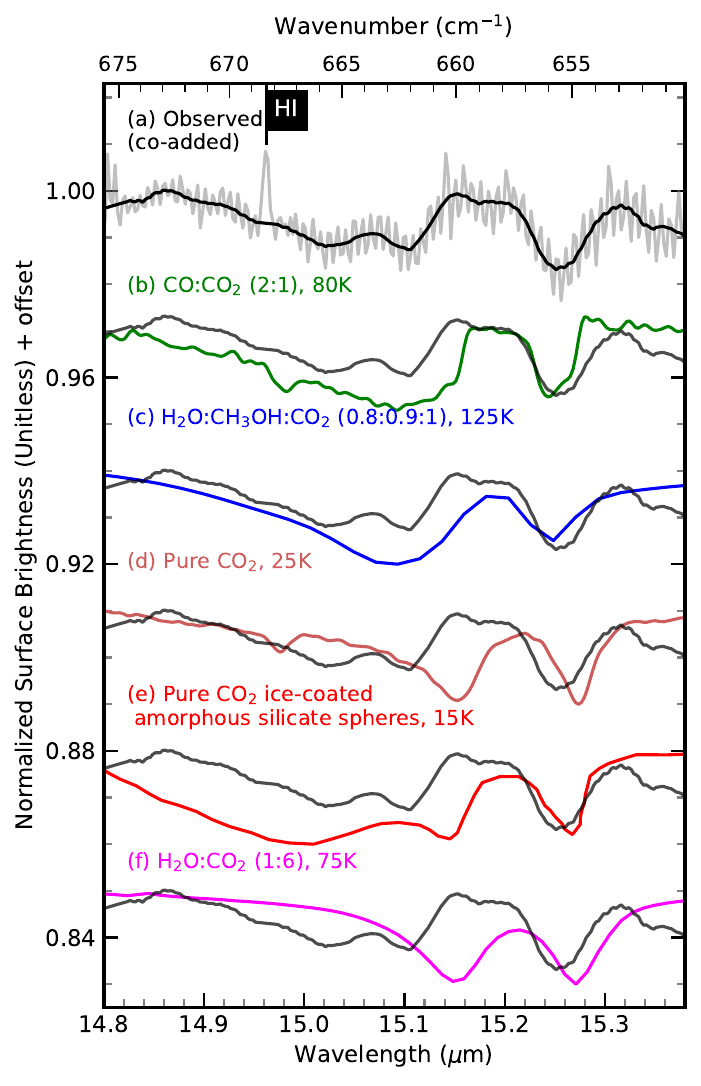}
     \caption{Comparison of observational and laboratory ice spectra. 
    \textit{(a)} Co-added \cotwo absorption profile across all pixels 
    with detectable \cotwo (gray), with the smoothed profile after 
    masking the \ion{H}{I} line at 14.96~\micron overplotted in black. 
    Laboratory spectra: \textit{(b)} CO:CO$_2$ (2:1) at 80~K (green); 
    \textit{(c)} \ce{H2O}:\ce{CH3OH}:\ce{CO2} (0.8:0.9:1) at 125~K (blue); 
    \textit{(d)} pure \cotwo at 25~K (brown); 
    \textit{(e)} pure \cotwo on amorphous silicate spheres at 15~K (red); 
    \textit{(f)} \ce{H2O}:\ce{CO2} (1:6) at 75~K (pink).}
    \label{fig:iceprofiles}
    
\end{figure}

The cold gas-phase \cotwo is located in the dusty torus of NGC~6302; this co-existence of dust and cold \cotwo gas mirrors conditions in envelopes of young stellar objects (YSOs) where \cotwo forms on cold dust grains, produces ice mantles, and desorbs when the temperature rises above the sublimation threshold. \cotwo ice formation thus appears plausible in NGC~6302 as well, and we therefore searched for the  characteristic $\nu_2$ bending mode at $\sim$15.2~\micron in the MIRI data.

To isolate potential ice absorption, we first divided the observations by the best-fit \cotwo gas model (see Fig.~\ref{fig:spectrum}). The resulting spectrum (brown curve in Fig.~\ref{fig:spectrum}) reveals a structured continuum containing spectral features from crystalline silicates \citep{2001A&A...372..165M,2002AandA...394..679K,2002Natur.415..295K}. This complexity introduces significant uncertainty in continuum placement. Comparison with laboratory spectra shows that clinoenstatite in particular bears similarities to the spectrum (see Fig.~\ref{fig:clinoenstatite}) and produces a strong absorption band at $\sim$15.4\micron that overlaps with the expected \cotwo ice band (Fig.~\ref{fig:spectrum}). Despite this contamination, we identify two key signatures of \cotwo ice: (1) a shallow, broad absorption between $\sim$14.9-15.15\micron, and (2) a second absorption between $\sim$15.2-15.3\micron. This characteristic double-peak structure matches laboratory \cotwo ice spectra and is clearly distinct from the single, deeper clinoenstatite band at 15.4 \micron or its weaker features. We fitted a local continuum (grey curve in Fig.~\ref{fig:spectrum}) to bracket these \cotwo features while avoiding regions dominated by silicate absorption. Given the dust complexity and continuum uncertainties, we cannot perform detailed ice mixture decomposition. However, the distinctive double-peak signature provides robust identification of \cotwo ice despite the challenging spectral environment.

To improve the signal-to-noise ratio, we co-added spectra from all spaxels containing gas-phase \cotwo (resulting in the grey curve in Fig.~\ref{fig:iceprofiles}a). The co-added spectrum reveals residual fringes that are not fully corrected by the MIRI data reduction pipeline. We applied a Savitzky-Golay filter 
\citep{1964AnaCh..36.1627S, 2005SigPr..85.1429L} to smooth out these fringes (black curve in Fig.~\ref{fig:iceprofiles}a). The resulting co-added spectrum reveals a clear double-peaked feature where $\sim$15.05~\micron feature is further split into two components and shows a broad blue wing.

Analysis of \cotwo ice absorption profiles in YSOs typically involves fitting a linear combination of five unique components to the data, each attributed to \cotwo in a different chemical environment \citep{2008ApJ...678.1005P, 2024A&A...692A.163B, 2025ESC.....9.1992B, 2025A&A...697A..53P}. However, this approach is not suitable for our data. Individual spaxels lack sufficient S/N for reliable component decomposition, while the co-added spectrum represents a spatial average over disparate sightlines with varying physical conditions, making the interpretation of fitted components ambiguous. Given the continuum uncertainties described above and possible contamination from dust features, we instead rely on direct comparison with laboratory spectra to identify the presence of \cotwo ice.

We compared the observed 15.2~\micron absorption profile with laboratory spectra from the Leiden Ice Database for Astrochemistry \footnote{\url{https://icedb.strw.leidenuniv.nl/spectrum_data}}, including pure \cotwo ice and mixtures with CO, H$_2$O, CH$_3$OH, and other species at various temperatures. The observed double-peak structure best resembles CO:CO$_2$ (2:1) at 80~K and H$_2$O:CH$_3$OH:CO$_2$ (0.8:0.9:1) at 120~K \citep[Fig.\ref{fig:iceprofiles}][]{1999A&A...350..240E, 2006A&A...451..723V}, reproducing the characteristic peak positions and relative intensities. Spectra of pure \cotwo ice and H$_2$O:\cotwo mixture also show a double peak feature but lack a broad blue wing (at $\sim$15.15~\micron). A model of pure \cotwo-ice coated amorphous silicate spheres \citep{2013ApJ...766..117P} better resembles our profile with its triple-peak structure and broad blue wing. While \citet{2013ApJ...766..117P} presented a model at 15~K only, their tabulated peak positions show decreasing peak separation with increasing temperature, suggesting that higher-temperature models would provide a better match to our observations. Given that the exact profile is sensitive to the grain shape, size distribution and ice composition, we cannot reliably determine the specific ice mixture or temperature from our data, but the 15.2~\micron feature confirms the presence of CO$_2$ ice in the torus of NGC~6302.

We calculated the average column density of \cotwo ice from the co-added and smoothed spectrum as $N_\text{ice} = \frac{\int \tau_\nu \, d\nu}{A}$ where $\int \tau_\nu d\nu$ is the integrated optical depth and $A$ is the band strength. We used a value of $A=1.6\times 10^{-17}$ cm molecule$^{-1}$ \citep{2015MNRAS.451.2145B}, and derive \mbox{N$_{\rm ice}$ $\approx$ (1 $\pm$ 0.3) $\times$ 10$^{16}$ cm$^{-2}$}. The uncertainty here accounts for observational error.

\section{Discussion}
\label{sec:discussion}

Molecular ices (such as \water, CO, \cotwo, CH$_4$) are abundant in cold, shielded environments including dense molecular clouds, envelopes of YSOs, and protoplanetary disks \citep{2015ARA&A..53..541B, 2024ARA&A..62..243C}. Conditions required for the formation of \water ice are also met in dense AGB and post-AGB outflows as suggested by observations \citep{1988ApJ...334..209S, 1990ApJ...355L..27O, 1999A&A...352..587S, 2002A&A...389..547H, 2003ApJ...586L..81S, 2011AJ....141...80M}. However, PNe present much harsher conditions due to the intense UV fields from the central star, making ice formation or survival generally infeasible \citep{2006A&A...449.1101D}. Indeed, only three PNe show H$_2$O ice: CPD-56°8032 \citep{1999ApJ...513L.135C}, NGC 6537 \citep{2002A&A...382..184M}, and NGC 6302 \citep{2001A&A...372..165M}, all characterized by massive dusty tori providing exceptional shielding. 

Our detection of \cotwo ice in NGC~6302 represents the first identification of an ice species more volatile than H$_2$O in any PN. Ice mantle formation depends on the balance between condensation and desorption, which varies with the binding energy of involved species, grain surface composition, and local physical conditions. For example, H$_2$O (the most abundant astrophysical ice) has highest binding energy and can typically remain in solid state at temperatures up to 150~K. On the other hand, \cotwo and CO have lower binding energies and thus desorb (in pure form) at 70-90~K and 20-30~K, respectively \citep{2022ESC.....6..597M}. The detection of \cotwo ice therefore indicates both sufficient shielding and relatively cold temperatures in NGC 6302's torus.

JWST's high spatial resolution revealed that extinction toward the NGC~6302 torus ($A_V > 76$~mag; \citealt{2025MNRAS.542.1287M}) is an order of magnitude higher than previously inferred from optical/near-IR observations ($A_V \approx 6-8$~mag; \citealt{2005MNRAS.359..383M, 2011MNRAS.418..370W}). This extreme shielding protects the ice from the central star's intense UV field (T$_{\rm _{eff}}$ $\approx$ 220,000~K), allowing \cotwo ice to survive. While NGC~6302 may represent an extreme case, dense dusty tori are common around evolved stars, particularly in binary systems \citep[][and references therein]{2023MNRAS.521...35I}, suggesting molecular ices may be more prevalent in these environments than previously recognized. JWST/MIRI's spatially resolved spectroscopy can now identify such highly shielded regions and assess the prevalence of ices in other PNe.

The observed \cotwo ice profile exhibits double-peak structure, a characteristic of pure, crystalline \cotwo ice, suggesting that \cotwo is present either in the pure form or in a mixture that is thermally processed \citep[see Fig.~\ref{fig:iceprofiles};][]{1997A&A...328..649E, 1998A&A...339L..17E, 2006A&A...451..723V, 2013ApJ...766..117P}. The \cotwo gas-to-ice ratio $(N_{\rm gas}/N_{\rm ice})$ in NGC~6302 is $\sim$1, much higher than in YSOs \citep[$\sim$0.02--0.15;][]{2009ApJ...702L.128A, 2011ApJ...736..133A}, indicating a fundamentally different formation pathway in PNe environments. \cotwo may have formed primarily in the gas phase during AGB/post-AGB phase \citep[see e.g.][]{1998A&A...330L..17J, Cami:EPAqr, Malek:HR4049_midIR} with subsequent freeze-out onto grains. While models of AGB outflows have successfully reproduced observed \water ice column densities by incorporating accretion of gas-phase \water by gas-grain collisions \citep{2003A&A...401..599D, 2019MNRAS.490.2023V}, extending such models to include freeze-out of \cotwo in PN environments represents an important avenue for future work.

The presence of \cotwo ice in the PNe environment has important implications for ISM enrichment. \cotwo ice facilitates the surface chemistry required for the formation of complex organic molecules such as formic acid \ce{HCOOH}, glycolaldehyde \ce{CH2OHCHO}, and acetaldehyde
\ce{CH3CHO} \citep{2009ARA&A..47..427H, 2020ApJS..249...26J, Potapov2021, 2022ApJS..259....1G}. Once formed, UV processing of these ices can drive further chemical complexity and controlled release back to the gas-phase, suggesting that PNe may contribute to the interstellar reservoir of organic material. If these species survive the prevailing physical and chemical processing, they may ultimately be incorporated into new star-forming systems. Such chemistry must be included in current PN chemical models and ISM enrichment models.

To comprehend the full picture, characterizing the full ice inventory in NGC~6302 is critical. The most abundant ice species in astrophysical environments -- \ce{H2O}, \ce{CO2}, \ce{CO}, \ce{CH3OH}, and \ce{NH3} -- have strong near-infrared absorption bands accessible to JWST NIRSpec, which can also detect their gas-phase counterparts, providing complementary information on sublimation processes and the ice-gas balance. High spatial resolution observations would constrain the chemical pathways, temperature structure, and ice processing mechanisms, establishing whether ice chemistry is common in dense PN tori.

\section{Conclusions}
\label{sec:conclusions}

We report the detection of \cotwo ice in a planetary nebula through spatially resolved JWST/MIRI observations of NGC~6302. This discovery demonstrates that molecular ices can form and survive in the heavily shielded regions of massive circumstellar tori, challenging the prevailing view that planetary nebulae are hostile to ice chemistry. We find \cotwo gas and ice present along the same line-of-sight. The gas-to-ice ratio differs markedly from that observed in young stellar objects, pointing to distinct ice formation or processing mechanisms in evolved stellar environments. Characterizing the full ice composition (\ce{H2O}, \ce{CO}, \ce{CH3OH}, \ce{NH3}) and spatial distribution through JWST NIRSpec observations will be essential for understanding the chemical complexity of these systems. This opens planetary nebulae as a frontier for ice chemistry. Ice surface chemistry—crucial for complex organic molecule formation—operates in these environments, and other massive dusty tori may harbour similar chemistry. Chemical models must incorporate ice-mediated pathways.

\begin{acknowledgements}

This work is based on observations made with the NASA/ESA/CSA James Webb Space Telescope. All of the data presented in this article were obtained from the Mikulski Archive for Space Telescopes (MAST) 
at the Space Telescope Science Institute. The data of this specific observing program can be accessed via \href{http://archive.stsci.edu/doi/resolve/resolve.html?doi=10.17909/s1rn-1t84}{doi:10.17909/s1rn-1t84}. This study is based on the international consortium of ESSENcE (Evolved Stars and their Nebulae in the JWST era). This article/publication is based upon work from COST Action NanoSpace, CA21126, supported by COST (European Cooperation in Science and Technology).

C.B., S.C., J.C., E.P. and N.C. acknowledge support from the University of Western Ontario, the Canadian Space Agency (CSA) [22JWGO1-22], and the Natural Sciences and Engineering Research Council of Canada. K.E.K. acknowledges support from grant JWST-GO-01742.010-A and H.L.D. from JWST-GO-01742.004 A. FK acknowledges support from the Spanish Ministry of Science, Innovation and Universities, under grant number PID2023-149918NB-I00.
This work was also partly supported by the Spanish program Unidad de Excelencia
María de Maeztu CEX2020-001058-M, financed by MCIN/AEI/10.13039/501100011033.
\end{acknowledgements}

\bibliographystyle{aa}
\bibliography{ALL}

\begin{appendix}

\section{Column denisty and excitation temperature of gas-phase \texorpdfstring{\cotwo}{CO2}}

\begin{figure}[t]
    \centering
    \includegraphics[width=\linewidth]{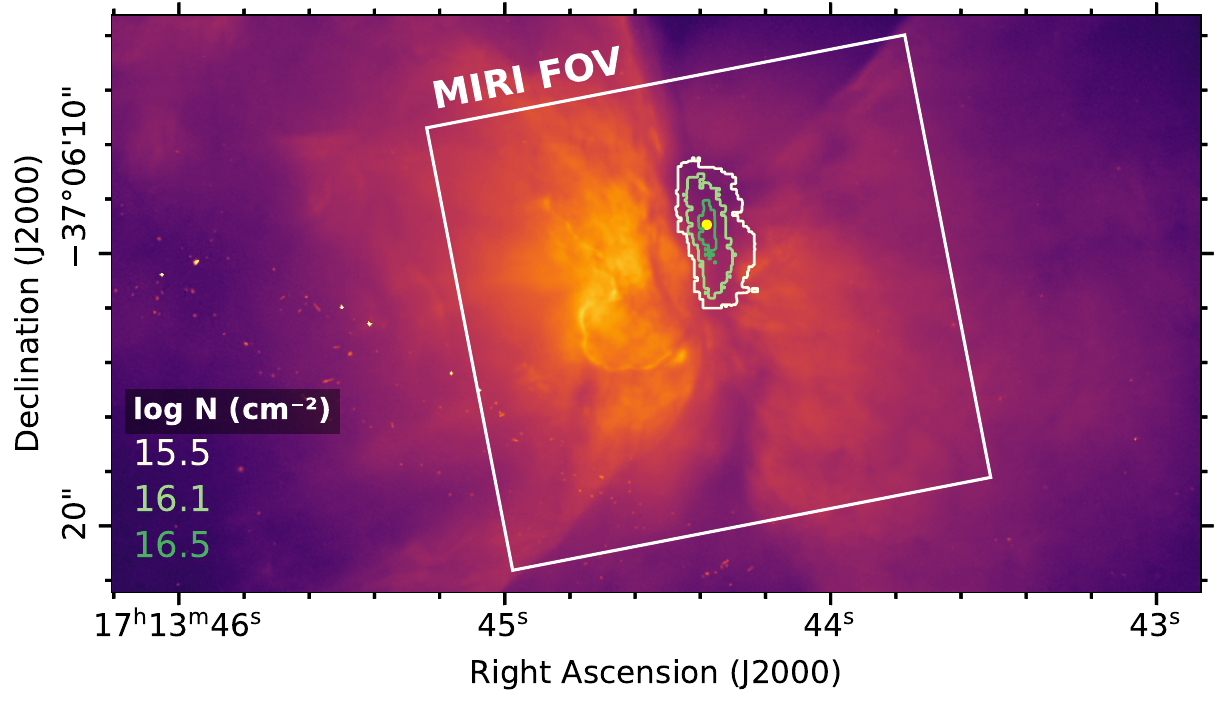}
    \caption{Location of \cotwo ice in NGC~6302. The image shows HST/WFC3 observations featuring filter F656N \citep{2022ApJ...927..100K}, which traces H$\alpha$ emission. The JWST MIRI mosaic is indicated by the white frame. Contours show the column density of gas-phase \cotwo, with corresponding log~N values (cm$^{-2}$) provided in the lower left. The yellow dot marks the pixel position  (R.A. $=$ 17$^h$:13$^m$:44.402$^s$, Dec. $=$ -37$^\circ$:06$'$:10.23 (J2000))used to extract the spectra shown in the Figs.~\ref{fig:spectrum}, \ref{fig:iceprofiles}, \ref{fig:12CO2_and_13CO2fit}, \ref{fig:clinoenstatite}. }
    \label{fig:HST}
\end{figure}

\begin{figure}
    \centering
    \includegraphics[width=\linewidth]{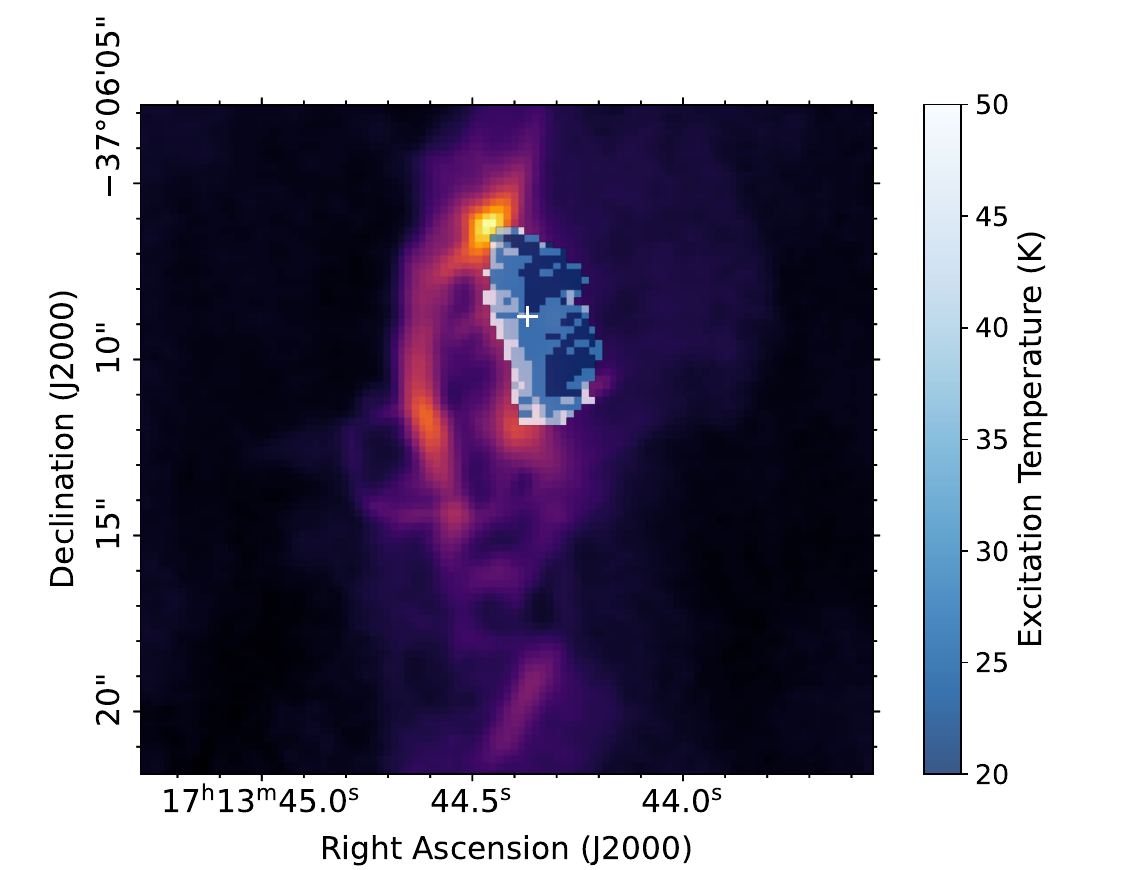}
    \caption{Spatial distribution of excitation temperature of \cotwo gas. The derived best-fit temperature of gas-phase \cotwo is mapped on the ALMA $^{12}$CO J$=$2--1 map \citep{2026ApJ...998....7M}. The temperature map reveals two blobs of colder \cotwo gas, which might indicate clumps in the torus, but needs further investigation.} 
    \label{fig:Tmap}
\end{figure}

Fig.~\ref{fig:HST} and Fig.~\ref{fig:Tmap} shows the spatial variations in column density and excitation temperature of gas-phase \cotwo. For details, see sec.~\ref{sec:gasCO2}. 

\section{Fitting \texorpdfstring{\twcotwo}{12CO2} and \texorpdfstring{\thcotwo}{13CO2}}
\label{app:12and13CO2}

\begin{figure*}[t]
    \centering
    \includegraphics[width = \textwidth]{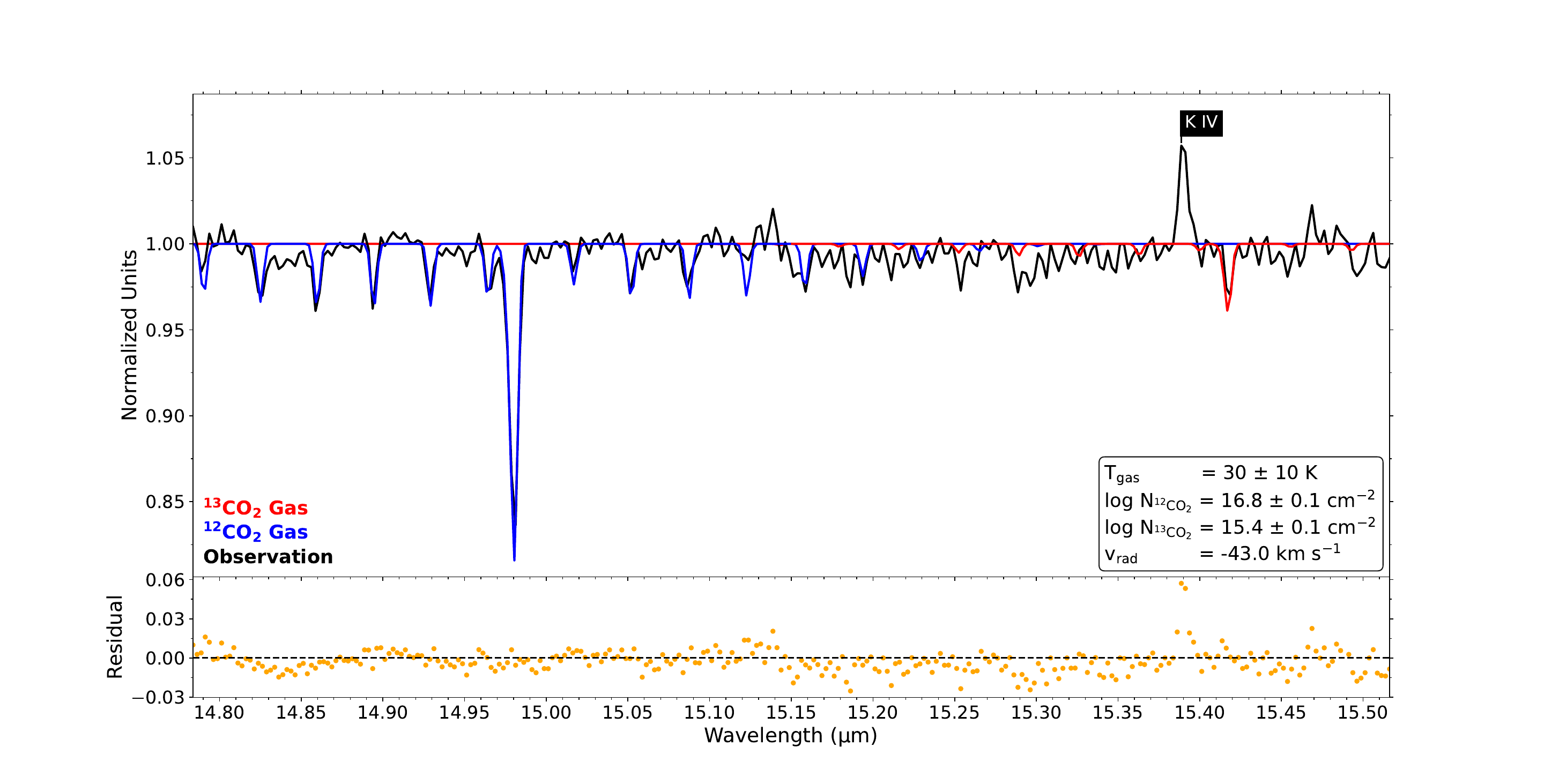}
    \caption{A representative fit of \twcotwo and \thcotwo. Normalized observed spectrum towards the northern part of the torus facing us (R.A. $=$ 17$^h$:13$^m$:44.402$^s$, Dec. $=$ -37$^\circ$:06$'$:10.23$''$ (J2000)) is shown in black and best-fit \twcotwo model and \thcotwo model are shown in blue and red, respectively.}
    \label{fig:12CO2_and_13CO2fit}
\end{figure*}

Along the same sightlines as \twcotwo, we detect much weaker spectral features due to gas-phase \thcotwo, which has a Q-branch at 15.42~\micron. We used the same fitting procedure as described in Sect.~\ref{sec:gasCO2} to determine the column density of \thcotwo, with excitation temperatures and radial velocities tied to the \twcotwo values for each pixel. An example fit is shown in Fig.~\ref{fig:12CO2_and_13CO2fit}. From these column densities, we derived \twcotwo/\thcotwo (and thus $^{12}$C/$^{13}$C) ratios in the range of 4--25, very similar to the range (4--15) reported by \citet{2026ApJ...998....7M} for NGC~6302 using ALMA observations for  $^{12}$CO/$^{13}$CO, H$^{12}$CO+/H$^{13}$CO+, and H$^{12}$CN/H$^{13}$CN. 

\section{Clinoenstatite absorption}

\begin{figure*}[t]
    \centering
    \includegraphics[width=\textwidth]{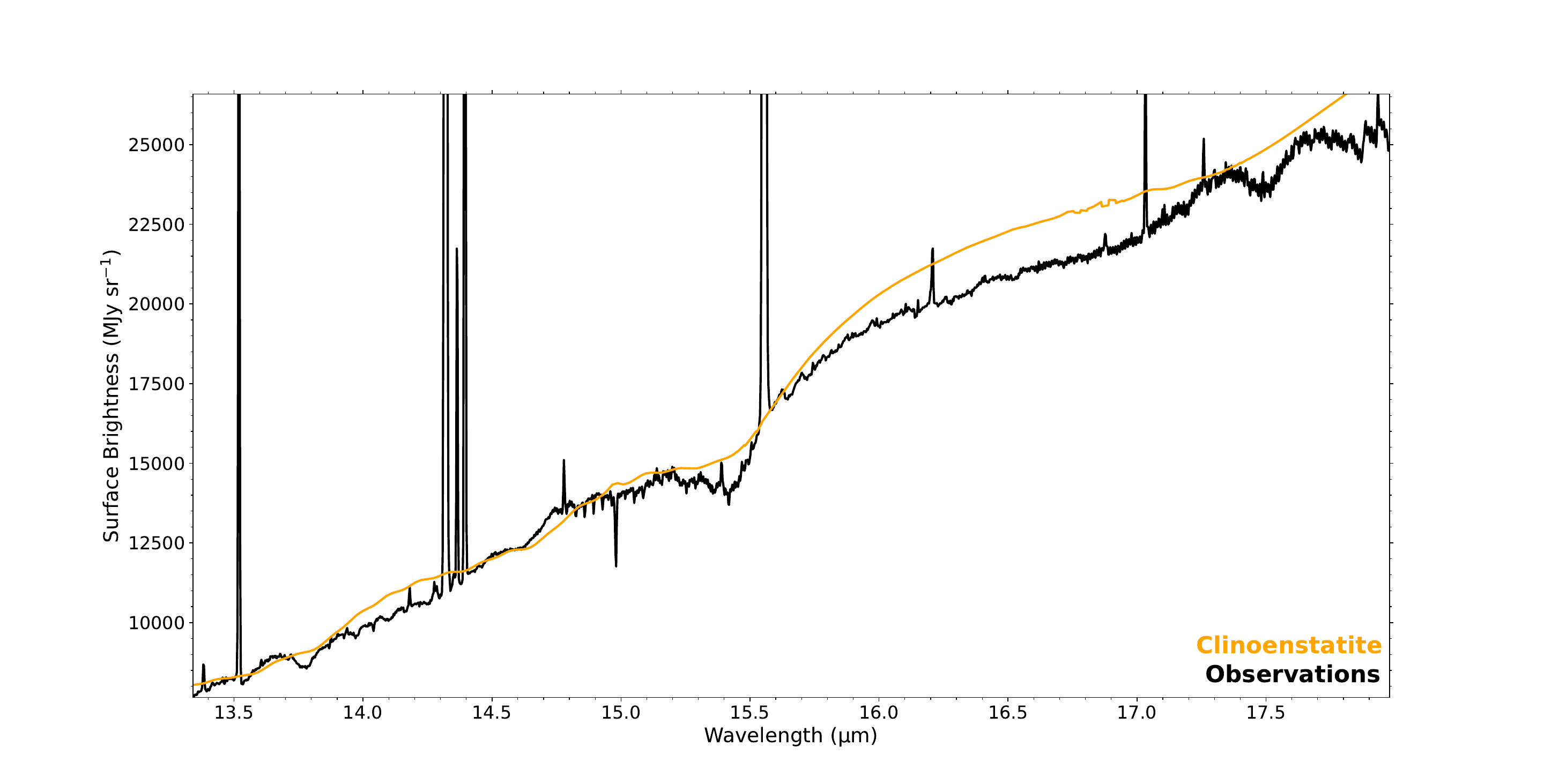}
    \caption{Comparison of observed spectrum with clino-enstatite model. The black curve represents the observed spectrum toward the same region as used to extract the spectrum in Fig.~\ref{fig:spectrum}. The orange curve shows the modelled spectrum of clino-enstatite, derived using its optical depth profile under the assumption of blackbody emission for both the background continuum and mineral component.  }

    \label{fig:clinoenstatite}
\end{figure*}

Fig.\ref{fig:clinoenstatite} compares the observed spectrum with the modelled spectrum of clino-enstatite. While this represents a simplified approximation of the dust continuum (which depends of many factors), clino-enstatite clearly cannot reproduce the absorption feature underlying the gas-phase \cotwo line at around 15.2\micron (see Sec.\ref{sec:ices}).

\end{appendix}
\end{document}